\documentclass[12pt,titlepage]{utarticle}

\newcommand{\smallhomoquot}[3]{#3\bigl\backslash\frac{#1}{#2}}
\newcommand{\vr}{\vec r}
\newcommand{\vo}{\vec \omega}
\newcommand{\vn}{\vec \nabla}
\DeclareMathSymbol{\square}{\mathord}{AMSa}{"03}
\DeclareMathSymbol{\rightsquigarrow}{\mathrel}{AMSa}{"20}

\newdimen\tableauside\tableauside=1.0ex
\newdimen\tableaurule\tableaurule=0.4pt
\newdimen\tableaustep
\def\phantomhrule#1{\hbox{\vbox to0pt{\hrule height\tableaurule width#1\vss}}}
\def\phantomvrule#1{\vbox{\hbox to0pt{\vrule width\tableaurule height#1\hss}}}
\def\ZZ{\relax{\sf Z\kern-.4em Z}}
\def\sqr{\vbox{%
  \phantomhrule\tableaustep
  \hbox{\phantomvrule\tableaustep\kern\tableaustep\phantomvrule\tableaustep}%
  \hbox{\vbox{\phantomhrule\tableauside}\kern-\tableaurule}}}
\def\squares#1{\hbox{\count0=#1\noindent\loop\sqr
  \advance\count0 by-1 \ifnum\count0>0\repeat}}
\def\tableau#1{\vcenter{\offinterlineskip
  \tableaustep=\tableauside\advance\tableaustep by-\tableaurule
  \kern\normallineskip\hbox
    {\kern\normallineskip\vbox
      {\gettableau#1 0 }%
     \kern\normallineskip\kern\tableaurule}%
  \kern\normallineskip\kern\tableaurule}}
\def\gettableau#1 {\ifnum#1=0\let\next=\null\else
  \squares{#1}\let\next=\gettableau\fi\next}

\input epsf
\newcount\figno
\def\fig#1#2#3{
\par\begingroup\parindent=0pt\leftskip=1cm\rightskip=1cm\parindent=0pt
\baselineskip=11pt
\global\advance\figno by 1
\epsfxsize=#3
\centerline{\epsfbox{#2}}
\vskip 12pt
{\bf Figure \the\figno:} #1\par
\endgroup\par
}
\def\figlabel#1{\xdef#1{\the\figno}}
\def\encadremath#1{\vbox{\hrule\hbox{\vrule\kern8pt\vbox{\kern8pt
\hbox{$\displaystyle #1$}\kern8pt}
\kern8pt\vrule}\hrule}}

\begin{document}

\preprint{
 HU-EP-97/44\\
 {\tt hep-th/9707259}\\
}

\title{Matrix Description of M-theory on $T^6$}
\author{Ilka Brunner and Andreas Karch
 \oneaddress{
  \\
  Humboldt-Universit\"at zu Berlin\\
  Institut f\"ur Physik\\
  Invalidenstra{\ss}e 110\\
  D-10115 Berlin, Germany\\
  {~}\\
  \email{brunner@qft1.physik.hu-berlin.de\\
  karch@qft1.physik.hu-berlin.de}
 }
}
\date{July 31, 1997}

\Abstract{We give some evidence that the worldvolume theory of
the M-theory KK 6-brane is governed by a non-critical membrane
theory. We use this theory to give a matrix description
of M-theory on $T^6$.}

\maketitle

\section{Introduction}

Recently there has been much excitement about a new approach to
capture the non-perturbative aspects of supersting theory: 
matrix theory (\cite{BFSS}). ``Matrix theory is a nonperturbative
Hamiltonian formalism for the theory formerly known
as string theory/M theory'' (\cite{banks}).
Nevertheless this nice new theory has a serious flaw:
it is formulated in the infinite momentum frame (IMF), sometimes
called lightcone gauge. Thus in addition to not being
manifestly covariant the theory becomes background dependent.
That means for every new background (e.g. for various compactifications)
one needs a different Hamiltonian, ergo a new theory. Since
we are mostly interested in compactifications down to four
dimensions (after all, this is the real world) it is
a very interesting problem to find the matrix description
of M-theory compactifications.
Since matrix theory is non-local, the compactified theory
has 'more' degrees of freedom than the theory describing
flat space.

Very early it has been realised that compactifications of
M-theory on tori can be realised by substituting the original
0+1 dimensional quantum mechanics by d+1 dimensional SYM with
16 supercharges (\cite{BFSS, Taylor}). This procedure
works only up to $T^3$. For higher dimensional tori the
new degrees of freedom added by going to higher dimensional SYM are
not 'enough'. This shows up as non-renormalizability of the
gauge theory. The UV of the SYM requires new degrees of freedom to
be well defined. The new degrees of freedom can not be determined
in any way from the IR information one has, so it is a difficult
task to find the matrix description for M-theory
on manifolds of dimension greater than 3.

Recently some progress has been made in this direction. Berkooz,
Rozali and Seiberg (\cite{rozber}) showed that the theory on M-theory
5 branes ( the (2,0) fixed point) can be used to define matrix theory
on the $T^4$. Similary one can obtain matrix theory on 4 manifolds
breaking higher amounts of SUSY (\cite{rozber2}). Later
Seiberg realised matrix theory on $T^5$ by some non-critical
string theory, the theory on NS 5-branes at zero string coupling
(\cite{seiberg}).
In this work we will argue
that the worldvolume theory on a KK 6-brane of M-theory is
described by a non-critical membrane theory which
can be used as the matrix description of M-theory on $T^6$.

While finishing this work we received \cite{arghh} where the authors
make a similar proposal.

\section{The KK-monopole}

Recently the KK monopole solution of higher dimensional SUGRA
theories have received renewed interest
\cite{Sorkin,Gross,Rub,Hull,Sen,Berg,Im,Sen2,Costa,Gibbons,Sen3}.
Classically these field configurations
appear after compactification of at least one direction.
They are the monopoles of the Kaluza Klein $U(1)$.
Since U-duality transforms these monopole 
solutions into the various D and NS branes, it is naturally
to also interpret them as some kind of brane, baptised KK brane.
Since the SUSY algebras in 11 and 10 dimensions naturally
contain the charges carried by these objects, it
has been speculated in \cite{Hull} that even the flat theory might allow
these KK $(D-5)$-brane ($D$ being the space-time dimension).

The multiple Kaluza-Klein monopole solution is described by the
metric:
$$ \label{e1}
ds^2 = -dt^2 + \sum_{m=5}^{D-1} dy^m dy^m + ds_{TN}^2\, ,
$$
where $D$ is 10 for various string theories and 11 for $M$-theory,
$y^m$ denote the space-like world-volume coordinates on the
$(D-5)$-brane represented by this solution,  and $ds^2_{TN}$ is the
metric of the Euclidean multi-centered
Taub-NUT space\cite{GIBHAW}:
$$ \label{e2}
ds_{TN}^2 = U^{-1} (dx^4 + \vec \omega\cdot d\vr)^2 + U d\vr^2\, .
$$
Here $x^4$ denotes the compact direction, and
$\vr \equiv(x^1, x^2, x^3)$
denotes the three spatial coordinates transverse to the brane.
$U$ and $\vo$ are defined as follows:
$$ \label{e3a}
U = 1 + \sum_{I=1}^N U_I, \qquad \vo = \sum_{I=1}^N \vo_I\, ,
$$
where,
$$ \label{e4}
U_I={\frac{4A }{ |\vr - \vr_I|}}\, ,
$$
and,
$$ \label{e5}
\vn\times \vo_I = \vn U_I\, .
$$
$A$ and $\vec r_I$ are parameters labelling the solution.
$\vec r_I$ can be interpreted as the locations of the
Kaluza-Klein monopoles in the transverse space.
In order that the solutions are
free from conical singularities at $\vec r=\vec r_I$, $x^4$
must have periodicity $16\pi A$.

The radius $A$ of the $U(1)$ part of the $SU(2) \times U(1)$ isometry
of the solution correponds to the compactification radius of
M-theory, which is associated to the coupling of the resulting IIA
string theory.

In what follows we will basically make use of the following
two results about the KK 6-brane of M-theory and its worldvolume
theory.

\begin{itemize}

\item The curvature of the monopole solution can be kept small
everywhere by taking the compactification radius big (see e.g \cite{Sen2}).
In the
limit $ A \rightarrow \infty$
\footnote{Remember that $A$ denotes the radius of the eleventh
dimension.
We would like to reserve the letter $R$ for the
light cone direction of matrix theory.} the curvature goes to zero
everywhere. Indeed it can be shown that
by making appropriate change of coordinates one
can map this to a flat metric (\cite{physrep}).
The KK brane decouples from gravity and since it is not
charged under the three form potential it decouples from
all the bulk M-theory fields.
The most obvious interpretation is that there is
just no brane left after decompactification.

\item For finite compactification radius M-theory on 
a circle can be interpreted as type IIA string theory.
The KK brane becomes the D6 brane. The effective
worldvolume theory is 6+1 dimensional $U(N)$ SYM,
where $N$ denotes the number of coincident 6-branes (\cite{Sen2,Hull,Berg}).
The
coupling can be determined using Polchinski's D-brane
results (\cite{Polchinski}) and turns out to be 
$$\frac{1}{g^2} = \frac{M_s^3}{g_s}
=\frac{1}{l_{p,11}^3}.$$
We used Wittens M-theory/IIA relations (\cite{M2A})
to express everything in terms
of $l_{p,11}$ and $A$, the radius of the 11th dimension,
since in the end we want to take the limit of flat 11 dimensions. 
We see that in these variables the coupling of the 6 brane
worldvolume theory is independend of the compactification radius 
(and this is only true for the 6 brane) \footnote{It is not clear to us
if just replacing type II parameters by 11d parameteres does indeed
capture the full low-energy physics even at large $A$. We
assume it does. The highly consistent picture that arises
hopefully helps to justify this assumption}.

\end{itemize}

Since the low-energy effective theory on the brane is totally
independent of the compactification radius and hence does not see the
flat limit as something special at all,
a different interpretation of this flat limit might be appropriate.
Let us briefly explain a proposal for the flat limit of the KK
monopole. Later we will work out consequences and show that
it leads to a highly consistent picture.

We take the fact that the low-energy effective theory on the
worldvolume is independend of the compactification radius
as an indication that in the decompactification limit the 6 brane does not
disappear, but decouples from the bulk fields.
That it has to decouple can be seen from the fact that 
the metric becomes flat and naively the brane disappears.
To an
11d observer away from the brane decoupling is just as
fine as disappearing.
He/She doesn't feel at all that there is a brane, it might as well be
gone (and hence this statement is consistent with the flat KK metric).
But
for the theory ON the brane it has consequences
\footnote{One might want to worry about the coupling of the excitations
on the brane to the bulk gravity. Take excitations on the KK-brane
of a given energy-scale $E$. Since they have to preserve the isometry
their energy density is constant along the circle and hence can be made
arbitrarily small everywhere by taking $A$ to be big enough. Since
gravity couples to energy density, these fluctuations also decouple.}.

The
theory on the brane stays well defined. It has a scale
$\frac{1}{l_{p,11}^3}$. Below this scale it is well described
by 6+1 dimensional SYM. At this scale new degrees of freedom
become important. For any finite compactification radius,
these degrees of freedom are described by the full fledged 11d theory.
In the flat 11d limit the new degrees of freedom that take over at the scale
SYM breaks down are no longer the full 11d quantum theory, since the bulk
fields (especially gravity) decoupled!

The conjecture thus is that in the flat 11d limit the KK 6-brane
is described by a consistent 7d quantum theory decoupled from gravity.

\section{Matrix Theory}

\subsection{The theory on the brane}

Given the existence of such a theory, a natural thing to do would
be to take its large $N$ limit, where $N$ denotes the number
of coincident KK 6-branes, as a matrix description (\cite{BFSS})
of M-theory on $T^6$. The
consistency of the emerging picture gives some evidence to our claims.
We thereby also get some further insight about the nature of
the theory on the brane.

This proposal is actually dual to Seibergs
proposal for matrix compactifications on the 6 torus \cite{seiberg}:
He takes the 5 brane of a type II string theory, say
IIB, at zero string coupling and takes a circle
transverse to the world volume to be compact. This
compact circle corresponds to one of the space-time
circles, the other 5 are related to the base-space
of the world volume theory on the 5 brane.
Now S-dualize this setup to get a D5 brane at infinite
coupling, T dualize the circle to get a IIA D6
brane at infinite coupling, now all compact
directions being worldvolume directions.
Now we can reinterpret the D6 brane at infinite
coupling as the KK brane in flat eleven dimensions.
In Seibergs analysis (\cite{seiberg}) the stringy excitations
of the system are described by the fundamental string stuck
to the NS5 brane. Under the above series of dualities
this F1 string turns into the M2 brane. 
We hence can make our proposal more concrete: The worldvolume
of the KK 6 brane in the flat 11d limit is governed by a supermembrane
theory living in 7 space-time dimensions. The low-energy
description of this theory is 6+1 dimensional SYM. This
theory indeed contains brane like excitations, which are the instantons
in 4 dimensions, become the Seiberg strings in 6 dimensions
and are membranes in 7 dimensions. Note also that the classical supermembrane
action is, like the Green-Schwarz string, only defined in certain
dimensions. 7 is one of them (\cite{wit}). In the recent paper of Aharony,
Berkooz, Kachru, Seiberg and Silverstein (\cite{Kachru})
they find a matrix description for this kind of theory.
They discuss a M5 brane with a compact
transverse 2 torus, which is, as described above, dual to what we 
are talking about.
They
remark that it is governed by a 2+1 dimensional field theory,
while Seiberg's stuck string is described by a 1+1 dimensional
field theory, which might be a further indication for
some membrane making its appearence.

\subsection{Quantitative Predictions}

Since we know that the low-energy description of this ficticious
7d-membrane theory is $U(N)$ SYM, we inherit the usual SYM on
the dual torus relations (\cite{BFSS, Taylor})
if we use it to do matrix theory.
Note that when people where using SYM on the dual torus
as the matrix description of M-theory on $T^6$ they
found quantitative agreement for all low-energy processes
(\cite{BFSS, rich}).
It is
thus a nice feature that our model reduces to SYM in the IR.
However it was noted very early that the SYM on the dual torus
description misses several degrees of freedom (\cite{shrink, rozali,
rozber, distler}). In field theory
this fact shows up as non-renormalizability. New degrees of
freedom become important in the UV. We will show in the following
that assuming that these additional degrees of freedom
are represented by the conjectured stuck membrane theory on
the KK worldvolume, one gets a correct description
of M-theory on $T^6$.

\subsubsection{The base-space/space-time relations}

We take the stuck-membrane theory living on a torus with
radii $\Sigma_i$ ($i=$1,...,6) and a membrane tension $\frac{1}{\lambda_p^3}$.
According to the usual terminology we will refer to this space
as the base-space. Certain properties of the base-space theory
can be obtained by looking at the way we constructed it as
the theory on a KK 6-brane in the flat limit. We will refer
to this as the auxiliary model. In the auxiliary model $\lambda_p$
becomes the 11-d Planck length. This base-space theory is supposed
to be the matrix description of space-time M-theory on $T^6$. This
space-time theory has a Planck scale $L_p$ (which is
not the same as $\lambda_p$) and radii $L_i$. According
to the usual SYM on the dual torus description we get the
following base-space/space-time relations (omitting factors of $2 \pi$):
\begin{eqnarray*} 
\Sigma_{1,2,3,4,5,6}^B &=& \frac{L_p^3 }{ R L_{1,2,3,4,5,6}} \\
\lambda_p^3=g^2 &=& \frac{L_p^{12}}{R^3 L_1L_2L_3L_4L_5L_6}
\end{eqnarray*}
where $R$ as usual denotes the light-cone radius of the matrix model
and $g$ is the gauge coupling of the SYM.

\subsubsection{Decompactification}

As first check let us show, that this proposal reduces correctly
to Seiberg's stuck string (\cite{seiberg}) under decompactification of a
space-time radius (which after all is the matrix description
of M-theory on $T^5$). That is we want to take a space-time length,
say $L_6$ to infinity. From this we see that $\Sigma_6$ goes to
zero, the tension of the membrane ($\frac{1}{\lambda_p^3}$) goes
to infinity while the tension of membranes wrapping $\Sigma_6$
(and hence yielding strings in the resulting 6 dimensional theory)
stays constant. We are thus left with a 6 dimensional string
theory with
\begin{eqnarray} 
\Sigma_{1,2,3,4,5}^B &=& \frac{L_p^3 }{ R L_{1,2,3,4,5}} \\
M_s^2= \frac{\lambda_{11}^3}{\Sigma_6} &=&
\frac{R^2 L_1L_2L_3L_4L_5 }{ L_p^9}
\end{eqnarray}
exactly as expected.

This can also be seen from the auxiliary construction. We look
at KK-6 branes
compactified on the base-space torus, when a circle of the
base-space torus shrinks.
M-theory on a vanishing circle is again type IIA, this time
at zero coupling. Since we this time compactified
a direction in the KK 6-brane worldvolume we
turn it into a IIA KK 5-brane at zero string coupling.
This is according to Sen (\cite{Sen2}) another
realization of Seibergs string theory, as we will
also discuss in the following section.
The base-space
to space-time relations are like those of the IIB
NS 5-brane at zero coupling 
which are again those of the usual SYM on the dual torus.
We hence get a consistent decompactification. Equivalently
one could undo the series of dualities described
previously to get back to type IIB NS 5-branes
at zero string coupling with
the radius of the transverse circle going to infinity, which
is exactly Seiberg's realization of this theory.

\subsubsection{The moduli space}

According to the general philosophy of matrix theory the
moduli space of the space-time theory appears as the
parameter space of possible base-space compactifications.
This parameter space can be obtained from the auxiliary model.
Here it is obvious what we want to do: we look at compactifications
of M-theory on a seven torus in the limit that one of the circles
becomes flat, which is of course the same as the moduli
space of M-theory compactifications on a 6 torus, that is
$$\smallhomoquot{E_{6(6)}}{Sp(4)}{\Gamma}$$
as expected. Note that this kind of seemingly circular logic is
similar to what Seiberg had to do (\cite{seiberg}).
We use information about
M-theory on $T^7$ to get information about matrix theory on $T^6$.
We have to do this since so far the only information we
have about the 7d theory is what we obtained from M-theory.
But at some point we should get a better understanding of this
seven dimensional theory without using M-theory. This should
be much easier than studying full-fledged M-theory, since we
are dealing with only 7 dimensions and decoupled gravity. Nevertheless
we can see from what we already know about M-theory and hence about
the non-critical membrane theory that this theory has the right
moduli space. Even though this seems somewhat trivial, note that
in the original form of Seiberg's conjecture for $T^6$ (the type II
5-brane with
a compact circle, \cite{seiberg})
it is far from obvious how this moduli space should
appear. In our formulation it is manifest.

\subsubsection{Excitations}

To study the possible excitations of the system, we would
like to analyze the energy of various branes stuck to the D6 brane
in our limit (decompactification of the 11th dimension).

That is, we consider a type IIA compactified on a torus
with right angles and no $B$ field present. The energy
of a configuration of $N$ D6 branes and n other branes
which are all of the same type
in the decompactification limit is given
by (\cite{seiberg})
(remember that to avoid confusion with $R$ we denote the radius of
the eleventh dimension of this auxiliary model by $A$):
$$E = \lim_{A \rightarrow \infty} \sqrt {
\left( N \Sigma_1 \Sigma_2 \Sigma_3 \Sigma_4 \Sigma_5 \Sigma_6 T_6 
\right)^2 + \left(n \Sigma_{i_1}...\Sigma_{i_p} T_{brane}
\right)^2} - N\Sigma_1 \Sigma_2 \Sigma_3 \Sigma_4 \Sigma_5 \Sigma_6 T_6 $$
where $T_6$ denotes the tension of the D6 and $T_{brane}$ the tension
of the other p-brane. We should express $T_6$ and $T_{brane}$ in 11d
quantities since we want to analyze the limit where
$A$ goes to infinity at fixed 11d Planck scale $\lambda_p$. In
this limit most branes become degenerate with the D6 brane, that
is for them the above expression vanishes. 
To see which branes give a finite contribution we calculated
the tensions of all IIA branes that can be completely wrapped on
the 6 torus (that is everything except for the D8).
\begin{center}
\begin{tabular}{|c||c|c|}
\hline
Brane&$\frac{1}{g^2}$&$T_{brane}$\\
\hline
D0& $\frac{l_s^3}{g_s}=\frac{\lambda_p^6}{A^3}$
&$\frac{1}{l_s g_s}=\frac{1}{A}$\\
\hline
D2& $\frac{l_s}{g_s}=\frac{\lambda_p^3}{A^2}$
&$\frac{1}{l_s^3 g_s}=\frac{1}{\lambda_p^3}$\\
\hline
D4& $\frac{1}{l_s g_s}=\frac{1}{A}$
&$\frac{1}{l_s^5 g_s}=\frac{A}{\lambda_p^6}$\\
\hline
D6& $\frac{1}{l_s^3 g_s}=\frac{1}{\lambda_p^3}$
&$\frac{1}{l_s^7 g_s}=\frac{A^2}{\lambda_p^9}$\\
\hline
F1&
&$\frac{1}{l_s^2}=\frac{A}{\lambda_p^3}$\\
\hline
NS5& $\frac{1}{l_s^2}=\frac{A}{\lambda_p^3}$
&$\frac{1}{l_s^6 g_s^2}=\frac{1}{\lambda_p^6}$\\
\hline
KK5&
&$\frac{1}{l_s^7 g_s}=\frac{\Sigma_i^2}{\lambda_p^9}$\\
\hline
\end{tabular}
\end{center}
As above $\frac{1}{g^2}$ denotes the worldvolume gauge coupling,
$A$ the radius of the 11th dimension, $l_s$ the string length,
$g_s$ the string coupling and $\lambda_p$ the 11d Planck length
of our auxiliary model. To convert to 11d units we used Witten's
formulas (\cite{M2A}): $l_s=\frac{\lambda_p^{3/2}}{A^{1/2}}$,
$g_s=\frac{A^{3/2}}{\lambda_p^{3/2}}$. To get the
tension of the KK5 one has to consider how this one arises.
We take a KK 6-brane of M-theory and wrap it around A. Its
worldvolume is hence transverse to one of the circles, say $\Sigma_i$.
It thus has the same tension as the D6 with the role of $A$
and $\Sigma_i$ interchanged. While $\Sigma_i$ now determines
the tension as indicated above, $A$ appears as one of the
wrapped circles in the energy formula.

Inspection of this table shows us that the above energy
formula gives us zero for most of the branes. The only branes
yielding a finite contribution are the F1, the D4 and the KK5.
The resulting energies are
\begin{eqnarray*}
E_i^E&=&\frac{n^2 \Sigma_i^2 \lambda_p^3 }{
2N \Sigma_1 \Sigma_2 \Sigma_3 \Sigma_4 \Sigma_5 \Sigma_6} \\
E_{ij}^M&=&\frac{n^2 
 \Sigma_1 \Sigma_2 \Sigma_3 \Sigma_4 \Sigma_5 \Sigma_6
 }{ 2N \lambda_p^3 (\Sigma_i \Sigma_j)^2} \\
E^i&=&\frac{n^2 \Sigma_i^2 
 \Sigma_1 \Sigma_2 \Sigma_3 \Sigma_4 \Sigma_5 \Sigma_6 
}{ 2N \lambda_p^9} \\
\end{eqnarray*}
The indices E and M denote that these energies are seen as
electric and magnetic fluxes in the low-energy SYM. The
low-energy description misses the states corresponding to
the KK5. These excitations naturally combine themselves
into a 27 of the discrete $E_{6(6)}$ duality group.
Using the base-space/space-time relations and converting
to space-time masses one sees that these excitations correspond
to BPS particles with masses:
\begin{eqnarray*}
M_i^E&=& \frac{1}{L_i}\\
M_{ij}^M&=& \frac{L_i L_j}{L_p^3}\\
M_i&=& \frac{L_j L_k L_l L_m L_n}{ L_p^6} \\
\end{eqnarray*}
where $i,j,k,l,m,n$ are a permutation of 1,2,3,4,5,6.
These energies correspond exactly to those expected from
the M-theory compactification, 6 momentum modes
15 wrapped membranes and 6 wrapped fivebranes.

\subsubsection{The momentum multiplet}

In addition to these finite energy BPS states the matrix model also
has states which correspond in space-time to objects wrapping the
lightcone direction $R$. Since the matrix model procedure
\cite{BFSS} tells us to take $R$ to infinity, these states
are moved off to infinite energy in the end. Nevertheless
it was the membrane wrapping $R$ and one of the space-time
circles which allowed \cite{Susskind} to indentify
the base-space and space-time variables according
to what we called the usual SYM on the dual torus procedure.
In \cite{EGKR} the authors baptised this kind of
multiplet the momentum multiplet, since all the
states can be obtained by U-duality acting on base-space momentum
modes.

For the $d=6$ case we are interested in they find that the matrix description
should contain the following states:

\begin{center}
\begin{tabular}{|c|c|}
\hline
$E_{SYM}$&$M_{spacetime}$\\
\hline
$\frac{1}{\Sigma_i}$&$\frac{R R_i}{L_p^3}$\\
$\frac{\Sigma_i \Sigma_j}{\lambda_p^3}$&$\frac{R R_i R_j R_k R_l}{L_p^6}$\\
$\frac{\Sigma_i \Sigma_j \Sigma_k \Sigma_l \Sigma_m}{\lambda_p^6}$&
$\frac{V R_i R}{L_p^9}$\\
\hline
\end{tabular}
\end{center}
where $V= R_1 R_2 R_3 R_4 R_5 R_6$.
In space-time these objects obviously correspond to membranes, 5-branes
and 6-branes wrapping $R$ (the 6 brane wraps $R$ and 5 of the space-time
circles, its tension is given by $\frac{R_i^2}{L_p^9}$).
Together with the 27 BPS excitations
from above, the momentum mode around $R$ and the 6-brane with tension
$\frac{R}{L_p^9}$ wrapping all 6 space-time circles, these states
form a 56 of $E_7$, which is the full expected U-duality
group for finite $R$, where this matrix model really describes
M-theory on $T^7$. A related discussion can be found in \cite{hacver}.

\cite{EGKR} were able to identify the first two entries in the low-energy
SYM description of the matrix model. The first correspond
to momentum modes along the compact base-space directions. The second
one is the wrapped Yang-Mills instanton, which is a $(d-4)$ brane
for SYM on $T^d$. In our case it is just our non-critical membrane!
This is possible since in the flat $A \rightarrow \infty$ limit
we were taking to decouple gravity, the membrane tension is
independent of $A$ and hence the wrapped membrane yields a finite energy
state (for finite $R$). This in fact allowed us to treat the membrane
as the fundamental excitation of the full 7d theory. If we now ask
ourselves which other objects have finite tension in the $A \rightarrow \infty$ limit, a short inspection of the table with the tensions we worked out
previoulsy shows that there is precisely one more such object, the NS 5-brane
with a tension of $\frac{1}{\lambda_p^6}$. The wrapped 5 brane
gives us the 6 missing states with the right energy! 
All these objects transform into each other under U-dualities.
The fundamental U-duality invariant object \cite{EGKR} identified
as $\frac{\Sigma_1 \Sigma_2 \Sigma_3 \Sigma_4 \Sigma_5 \Sigma_6}
{\lambda_p^9}$ can be identified as the product of 5-brane tension,
membrane tension and base-space volume.

\section{Remarks}

In this work we conjectured the existence of a non-critical
membrane theory living in 7 space-time dimensions. The
existence of such a theory would lead to a unification
of the variuos 6 dimensional non-critical string \cite{seiberg, verlinde}
theories
like M-theory did for the various string theories in 10
dimensions. At some points of its moduli space the
compactified membrane theory is best described by
some non-critical string theory. We already saw that
compactification on a circle gives us back Seiberg's
6 dimensional type II stuck strings. Similar one
would expect that compactification on 
an interval leads to the stuck type I.

If this story is right it would mean that to incorporate the 5 brane
in matrix theory, we should not only include the particles charged
electrically under the KK $U(1)$, which are the zero branes, but
also the dual objects charged magnetically under the $U(1)$,
the KK monopoles. This might be a sign of how the principle of duality
becomes manifested in the matrix theory.

There is an obvious generalization to the work done so far. If one
wants to consider M-theory compactifications on 6 manifolds preserving
less supersymmetry ($K3 \times T^2$ or $CY$), the matrix description
of this should obviously be the non-critical membrane theory
described above living on a correponding manifold, breaking the same
fraction of supersymmetry. 

Remains the question how to describe compactifications on even higher
dimensional tori. So far we managed to compactify down to 5 dimensions.
But we definitely want to understand compactifications to 4 dimensions!
There is one more Kaluza Klein like object in M-theory, the conjectured
M9 brane. We call it Kaluza Klein like (\cite{9707139}), since
it naively only appears in
compactifications 
(\cite{9601150}), like the KK monopole. For example
double dimensional reduction of this M9 brane is supposed to
yield the D8 brane of IIA. From this we see that
the worldvolume of this M9 brane is 10d SYM (\cite{Hull}).
Like for the KK 6-brane, the SUSY algebra of 11d SUGRA
contains a central charge for a 9-brane, so it was
speculated that there might actually be a 9-brane solution
in uncompactified M-theory. Since the D8 only
is possible in a massive IIA background and there is
no known generalization of the massive type IIA to eleven dimensions,
it is not even possible to construct the 9-brane as
a classical SUGRA solution. 

After the experiences we made with the KK 6-brane one can nevertheless
hope for a similar mechanism taking place. Instead of disappearing
the worldvolume theory of the M9 brane might 
decouple from the bulk
fields in the flat 11d limit, giving rise to a well defined
10 dimensional theory without gravity. The gauge
coupling of the SYM on this hypothetical brane
can be obtained from the gauge coupling of the D8 which is:
$$\frac{1}{g_8^2}=\frac{1}{l_s^5 g_s}=\frac{A}{l_p^6}.$$
Since the SYM on the M9 brane should reduce to this after dimensional
reduction on the circle with radius A, its coupling would be
$$\frac{1}{g_9^2}=\frac{1}{g_8^2 A}
 =\frac{1}{l_p^6}$$
again independent of the radius of the eleventh dimension!
If existent this theory could help us to formulate the
matrix theory of M-theory compactifications on higher tori.

\section*{Acknowledgements}
We would like to thank B. Andreas,
G. Curio, H. Dorn, S. Govindarajan,
D. L\"ust, T. Mohaupt, B. Ovrut, C. Preitschopf, W. Sabra,
D. Waldram
and especially J. Distler, A. Hanany and A. Sen
for useful discussions. The work of
both of us is supported by the DFG.


\begin{thebibliography}{99}
\bibitem{Sorkin}
R. Sorokin, {\it Phys. Rev. Lett.} {\bf 51} (1983) 87.
\bibitem{Gross}
D. Gross and M. Perry, {\it Nucl. Phys.} {\bf B226} (1983) 29.
\bibitem{Rub}
P. Ruback, {\it Math. Phys.} {\bf 107} (1986).
\bibitem{Hull}
C. Hull, hep-th/9705162.
\bibitem{Sen}
A. Sen, hep-th/9705212.
\bibitem{Berg}
E. Bergshoeff, B. Janssen and T. Ortin, hep-th/9706117;
E. Bergshoeff, M. de Roo, E. Eyras, B. Janssen, J. P. van der Schaar,
hep-th/9704120.
\bibitem{Im}
Y. Imamura, hep-th/9706144.
\bibitem{Sen2}
A. Sen, hep-th/0707042.
\bibitem{Costa}
M. Costa and G. Papadopoulos, hep-th/9612204.
\bibitem{Gibbons}
G. Gibbons, G. Papadopoulos and K. Stelle, hep-th/9706207.
\bibitem{Sen3}
A. Sen, hep-th/9707123.
\bibitem{seiberg}
N. Seiberg, hep-th/9705221.
\bibitem{GIBHAW}
S. Hawking, {\it Phys. Lett.} {\bf 60A} (1977) 81;
G. Gibbons and S. Hawking, {\it Comm. Math. Phys.} {\bf 66} (1979) 291.
\bibitem{physrep}
Eguchi, Gilkey and Hanson, {\it Phys. Rep.} {\bf 66} 363.
\bibitem{Polchinski}
J. Polchinski, hep-th/9611050 and references therein.
\bibitem{M2A}
P. Townsend, {\it Phys. Lett.} {\bf B350} (1995) 184, hep-th/9501068;
E. Witten, {\it Nucl. Phys.} {bf B443} (1995) 85, hep-th/9503124.
\bibitem{BFSS}
T. Banks, W. Fischler, S. H. Shenker, L. Susskind, {\it Phys.Rev}
{\bf D55} (1997) 5112, hep-th/9610043.
\bibitem{wit}
see e.g. B. de Wit, hep-th/9701169.
\bibitem{Kachru}
O. Aharony, M. Berkooz, S. Kachru, N. Seiberg, E. Silverstein, hep-th/9707079.
\bibitem{Taylor}
W. Taylor, {\it Phys. Lett.} {\bf B394} (1997) 283, hep-th/9611042.
\bibitem{rich}
David Berenstein, Richard Corrado, hep-th/9702108.
\bibitem{distler}
David Berenstein, Richard Corrado, Jacques Distler, hep-th/9704087.
\bibitem{shrink}
W. Fischler, E. Halyo, A. Rajaraman, L. Susskind, hep-th/9703102.
\bibitem{rozali}
Moshe Rozali, {\it Phys. Lett.} {\bf B400 } (1997) 260, hep-th/9702136.
\bibitem{rozber}
Micha Berkooz, Moshe Rozali, Nathan Seiberg, hep-th/9704089.
\bibitem{rozber2}
Micha Berkooz, Moshe Rozali, hep-th/9705175;
Suresh Govindarajan, hep-th/9705113.
\bibitem{Susskind}
Leonard Susskind, hep-th/9611164.
\bibitem{EGKR}
S. Elitzur, A. Giveon, D. Kutasov, E. Rabinovici, hep-th/9707217.
\bibitem{hacver}
Feike Hacquebord, Herman Verlinde, hep-th/9707179.
\bibitem{verlinde}
R. Dijkgraaf, E. Verlinde, H. Verlinde, hep-th/9704018;
\bibitem{banks}
for a recent review see: T. Banks, hep-th/9706168
\bibitem{arghh}
A. Losev, G. Moore, S. Shatashvili, hep-th/9707250.
\bibitem{9707139}
P. S. Howe, N. D. Lambert, P. C. West, hep-th/9707139.
\bibitem{9601150}
E. Bergshoeff, M. de Roo, M. B. Green, G. Papadopoulos, P. K. Townsend,
hep-th/9601150
\end{thebibliography}
\end{document}